# Looking At Situationally-Induced Impairments And Disabilities (SIIDs) With People With Cognitive Brain Injury

**Osian Smith**
869024@swansea.ac.uk
Swansea University
Swansea, United Kingdom

**Stephen Lindsay**
s.c.lindsay@swansea.ac.uk
Swansea University
Swansea, United Kingdom

## ABSTRACT

In this document, we discuss our work into a speaker recognition to support people with prosopagnosia and the limitations of alerting the user of whom they are in discussion with. We will discuss how current research into Situationally Induced Impairments Disabilities (SIIDs) can assist people with disabilities and vice versa and how our work can support people who may find themselves in a situation where they are impaired with facial recognition.

**INTRODUCTION**

Situationally-induced impairments and disabilities (SIIDs) are an expanding issue in ubiquitous computing. While our research area is supporting patients with Cognitive Brain Injury (CBI), the work of SIIDs supports our research interests by allowing existing devices to be more accessible for people with impairment. Current estimates of the abandonment rate Adaptive Tools (AT) is as high as 75%, with many of people using AT finding that they give themselves stigma of being unwell [13]. By using standard technologies, we can remove the stigma of the patient using a medical device, which in turn can increase the life quality of the patient.

We chosen to focus on supporting patients with prosopagnosia by speaker recognition on a smartwatch as we believe that by using a smartwatch, the patient should not feel any stigma as smartwatches are commonplace. We also feel that speaker recongition can support people with SIIDs.

**PROSOPAGNOSIA**

Prosopagnosia, commonly known as facial blindness, is the condition where patients cannot recognise faces, people or voices [11]. Prosopagnosia is estimated to affect around 2.5% of the population, approximately the equivalent number of people who have dyslexia [8]. It can be cognitive at birth or acquired through CBI and although treatment is available, it is widely accepted not all will be successful with many patients with no suitable treatment [4, 6, 14].

Prosopagnosia is usually present in an individual at birth although can be acquired. There is evidence that suggests that prosopagnosia can be a trait of an individual with Autism Spectrum Disorder, although this link cannot be considered established due to little evidence supporting them. There is also debate whether prosopagnosia should be kept as its condition when ASD cannot be used to explain face blindness [11, 22]. People who suffered brain trauma can also develop prosopagnosia due to damage to the segment of the brain that processes facial data. Prosopagnosia can also be genetic with evidence showing that the ability to recognise faces is an evolutionary trait as research shows that the Macaque Monkey has neurons specifically tuned to facial recognition [10, 11, 18].

There is no treatment for prosopagnosia that has demonstrated an improvement to all patients with many patients demonstrating no improvement to to treatment[2, 14] . Treatments for prosopagnosia

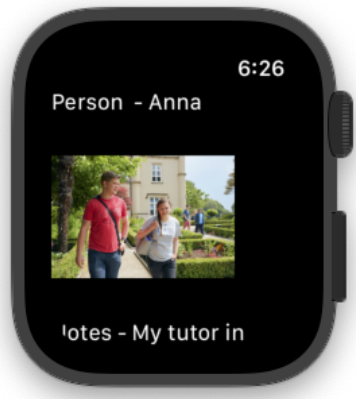

**Figure 1: A screenshot of our prototype application displaying information on a speaker after identifying an individual. On this screen we display the person's name, a photo and notes on them**

focuses on training of processing facial features [1, 3, 9, 17]. This adds a cognitive load to the patient which may be unstable for people with ABI due to their decrease cognitive ability [2, 6]

**OUR CURRENT WORK**

Our current research is investigating supporting patients with prosopagnosia to identify the person that the patient is in conversation. We are currently developing on a system that uses a microphone on a smartwatch to then listen to speakers, before sending the audio to the smartphone for further processing. The smartwatch then uses a one-shot learning neural network to identify the person in conversation with on their smartphone before alerting them on the smartwatch.

We are currently running participatory design workshops with patients with prosopagnosia to understand how they would want such a device to work. Users did not want to have a device that showed that they had a condition, such as a microphone on their neck but would prefer a smartwatch like the Apple Watch, a device that many people already wear. However, many of the participants do have impairments that may make it difficult to use the smartwatch.

As a result, we are using these participatory design workshops to understand the impairments of using a device. For example, we are looking into how the user can interact with their smartwatch to activate the microphone without making it evident to the person who they're in converse. This research may also benefit the SIIDs community, by understanding how to interact with a smartwatch without people knowing - such as sending a message to a family member while the user is in a meeting.

We are also interested in alerting the user that we have identified the speaker. Our current working prototype as seen in Figure 1 uses visual feedback of showing the user who the speaker is from a photo and a quick description. However, we are investigating audio feedback, where the user has an earphone in their ear which an automatic voice speaks to them stating who it is, or do we have the smartwatch give off a unique chime for every user. Could we also use haptic feedback, where we use the watch's vibration motor to give unique vibration to every person in the speaker?

There are limitations to each of these modalities. The chime and haptic feedback add a cognitive load onto the user, which may not be suitable for patients with CBI due to their diminished cognitive ability [2, 6]. However visual feedback may be rude because the user has to look at their smartwatch which may give the impression that the user is ignorant. Our headphone implementation could also be considered being rude as it may seem that the user is listening to music when in conversation.

Many of these concerns are also valid to the general population with SIIDs. How can users interact with their devices without being rude [7]? Is the user in a suitable situation where they can use voice assistance? Do these technologies cause concern for user and bystander privacy? Google Glass was infamous for the lack of bystander privacy with many of its users being called "glassholes" [5].

Furthermore, research by McNaney et al. found that during his study with Google Glass, many patients felt uncomfortable wearing this technology due to privacy concerns [12].

**Situationally Induced Impairments Disabilities**

Many patients who require AT can also benefit from the research into SIIDs. For example, patients with Parkinson's disease may struggle to interact with a smartphone device due to their tremors [15]. Many users who don't have Parkinson's may find themselves in a situation that they cannot use a device for similar reasons of patients with Parkinson's, such as being in a moving vehicle. As a result, could an interface designed for someone moving in a car also be transferred to someone with Parkinson's?

Research by Trewin [20] has previously researched dynamic keyboards and how people with motor disabilities and SIIDS interact with their phone. Trewin noted that there were technologies such as sticky keys to help people use keyboards, however Trewin noted that their were difficulties with adjusting the delays which many users may not want to constantly change [20]. While people with motor disabilities would likely configure the settings correctly, there may need to be more of an adaptive approach to people with SIIDS. We do have to consider that Trewin works was published in 2004 before touchscreen smartphones became commonplace.

Research by Nicolau et al. [15] found that people with motor disabilities while can tap on keyboards, although their error rate was higher than participants with no impairments [15]. While technologies such as predictive text and dictation may help, we did not find any literature that investigate this area.

Wobbrock [21] believes that once SIIDs are better understood, research should focus on comparing SIIDs to medical conditions such as comparing people with poor eyesight to people with poor vision. From this we can gather relationships between conditions and have better design decisions that support a wider range of people, which would also reduce stimga in these technologies.

Research into the design process of applications by Tigwell et al. [19] found that many clients were resistant to accessibility as they felt that the designers were overbearing. Tigwell et al. stated that while they can use well-known guidelines such as they Apple Human Interface Guidelines [1] and Google Material Design [2], they may become problematic due to the amount of content available. There are also inconsistencies with these guidelines with Apple's low contrasting fonts being challenging for people with typical vision to read. [16] [19]

Tigwell et al. [19] made the argument that this approach to accessibility could be made to support people with SIIDs, although we believe that this argument of supporting SIIDS could also support users with impairments.

Our research on supporting patients with prosopagnosia can also be deployed to support patients with SIIDs. Many people without prosopagnosia find remembering names challenging, especially in certain situations, for example, a lecture at a university. A lecture may be responsible for 300+ students enrolled on a module and may need to discuss an important topic with one student; however, they cannot remember every student on the course. By using the speaker recognition system that we developed, could the lecture put notes on that student and be alerted when they are talking to the student to discuss the important topic?

---

[1] https://developer.apple.com/design/human-interface-guidelines/

[2] https://material.io/design/

## CONCLUSION

In this document, we have discussed the high abandonment rate of Assistive Technologies with research pointing this is a result of stigma. We have also outlined our work that supports patients with prosopagnosia by listening to the person they are in conversation with and stating who they are talking to. We also highlighted how research into SIIDs can help people who require AT and vice versa and that our work could also support people who do not suffer from prosopagnosia.